\documentclass[11pt]{article}
\begin{document}

\title{{\bf Properties of Moyal-Lax Representation}}
\author{Ashok Das\\
Department of Physics and Astronomy,\\
University of Rochester,\\
Rochester, NY 14627-0171\\
USA\\
\\
and\\
\\
Ziemowit Popowicz \\
Institute of Theoretical Physics, \\
University of Wroc\l aw,\\
50-205 Wroc\l aw\\ 
Poland.}
\date{}
\maketitle

\begin{abstract}

The properties of standard and the nonstandard Moyal-Lax representations are
systematically investigated.
It is shown that the Moyal-Lax equation can be interpreted as a
Hamiltonian equation and can be derived  
from an action. We show that the parameter of non-commutativity, in
this case, is related to  the central charge 
of the second Hamiltonian structure of the system. The Moyal-Lax description
leads in a natural manner  
to the dispersionless limit and provides the second  Hamiltonian
structure  of dispersionless integrable models, which has been an open
question for sometime.

\end{abstract}

\newpage

\section*{I. Introduction:}
Integrable models \cite{1}, both bosonic as well as supersymmetric
\cite{2},  have played important roles in the study of 
conformal field theories, strings, membranes as well as topological field theories. In recent years,
it has become known that string (membrane) theories naturally lead to non-commutative field theories \cite{3}, 
where usual multiplication of functions is replaced by the star
product due to Groenewold \cite{3'}, \cite{4}. It is interesting, therefore, 
to ask if integrable systems can also be described in terms of star
products and Moyal brackets \cite{4}.
It appears that there are two possible approaches to this problem. In
the  first method \cite{5} one can formulate the
soliton theory in the non-commutative space-time which is realized
using  the star product. In the second approach,
the star product can be used directly in the Lax operator
description \cite{5'}-\cite{6} or in the zero-curvature condition
\cite{6'}-\cite{7}.   

In this letter, we follow the second approach and  show that it is
possible  to use the  Moyal bracket as the Poisson bracket in the
phase  space.
Such an interpretation allows us to present a Moyal-Lax
representation for  the soliton system and study such 
representations systematically. So far, the Moyal bracket has been used to
construct only the standard Lax representation for bosonic integrable
systems \cite{5'}. We show that such brackets can be used to
describe nonstandard representations as well.
There are many interesting features
that emerge from such a representation 
and we argue that such a representation may, in fact, be more desirable. Among the various interesting features 
that emerge, we note that the parameter of non-commutativity, in such
an analysis,  is directly related to the central 
charge of the second Hamiltonian structure of the system. Furthermore, in such a formulation, the Lax equation has a natural 
interpretation of a Hamiltonian equation and can be simply derived from an action. Furthermore, such a description 
naturally  leads to the dispersionless limit (in which the models become related to membranes and topological field 
theories) and thereby allows us to derive the Hamiltonian structures
(first, second,...), which has been an open 
question for quite some time. 

\section*{II. Basic Definitions:}
Integrable systems are Hamiltonian systems and, therefore, are
naturally defined  on a phase space. The star product  
of two functions, on this space, is defined to be
\begin{equation}
 A(x,p) \star B(x,p) = e^{\kappa(\partial_x\partial_{\widetilde{p}} {~} - {~} \partial_p\partial_{\widetilde{x}})} 
A(x,p)B({\widetilde{x}},{\widetilde{p}} )  \Big | _ {{\widetilde{x}=x},{\widetilde{p} = p}}.
\end{equation}
The conventional Moyal bracket, then, follows to be 
\begin{equation}
 \Big \{ A(x,p) , B(x,p) \Big \}_{\kappa} = \frac{1}{2\kappa} \Big ( A \star B - B \star A \Big ). 
\end{equation} 
Here $\kappa$ is the parameter of non-commutativity, which, as we will see, is directly related to the central 
charge of the second Hamiltonian structure in the case of integrable models. From (1) and (2), it follows 
immediately that 
\begin{equation}
\lim_{\kappa \rightarrow 0}\; \Big \{ A,B \Big \}_{\kappa} =   \Big \{
A,B \Big \}, 
\end{equation}
where the last bracket in (3) stands for the standard canonical Poisson bracket.

The star product gives the momentum an operator character. In
particular,  let us note that for 
any arbitrary integer $m,n$ (positive or negative) 
\begin{eqnarray}
p^{n} \star p^{m} & = & p^{n+m} \nonumber\\
p^n \star f(x) &=& \sum_{m=0}  \left( \begin{array}{c} n\\m \end{array}  \right)(-2\kappa)^mf^{(m)} \star p^{n-m},
\end{eqnarray}
where 
\begin{equation}
\left( \begin{array}{c} n\\m \end{array} \right) = \frac{n(n-1)...(n-m+1)}{m!},{~~~}{~}
\left( \begin{array}{c} n\\0 \end{array} \right) =1, 
\end{equation}
and $f^{(m)}(x)$ stands for the $m$-th derivative of $f(x)$ with respect
to $x$. We note that these are precisely the relations (up to
normalizations) satisfied by the derivative operator.

With these, we can define two classes of Lax operators on the phase space as 
\begin{eqnarray}
L_n &=& p^n+u_1(x)\star p^{n-1} + u_2(x)\star p^{n-2} + ... + u_n(x) \nonumber \\
\Lambda_n &=& p^n + u_1(x)\star p^{n-1} + ... + u_n(x) +
u_{(n+1)}\star p^{-1} + ... ,
\end{eqnarray}
which we can identify respectively with the Lax operator for the
generalized KdV  hierarchy and the KP hierarchy [8]. In simple 
terms, we have replaced the space of pseudo-differential operators by
polynomials in momentum, which nonetheless inherit 
an operator structure through the star product and define an
algebra. We will call such an algebra  the 
Moyal momentum algebra ($Mm$ algebra)\footnote{ Notice that our Mm
algebra is different from  the concept of pseudo-differential 
operators with the coefficients taken from the Moyal algebra
introduced in \cite{9}.}.
It is easy to check that all the known properties of pseudo-differential 
operators carry through with suitable redefinitions. For example, for any two arbitrary operators $A$ and $B$,
which are elements of the $Mm$ algebra, the residue (the coefficient
of  the $p^{-1}$ term with respect to the Moyal product) of the Moyal
bracket can 
be checked to be a total derivative, namely 
\begin{equation}
Res \Big \{ A,B \Big \}_{\kappa} = ( \partial_{x} C).
\end{equation}
Consequently, one can define 
\begin{equation}
Tr A = \int dx {~} Res {~} A,
\end{equation}
which is unique (with the usual assumptions of asymptotic fall off)
and which satisfies cyclicity.  

For a general Lax operator $\Lambda_n$, it is straightforward to show that 
\begin{equation}
\frac{\partial \Lambda_n}{\partial t_k} = 
\Big \{\Lambda_n,  \Big (\Lambda_n^{\frac{k}{n}}\Big )_{\geq m}\Big \}_
{\kappa}, {~}{~}   k\neq ln,
\end{equation}
where $k,l$ are integers, defines a consistent Lax equation, provided $m=0,1,2$ and that the 
projections are defined with respect to the star product. Note here
that  $A^{\frac{k}{n}} = 
A^{\frac{1}{n}}\star A^{\frac{1}{n}} \star ... \star A^{\frac{1}{n}}$
involving $k$ such factors and the 
$n$-th root is determined formally in a recursive manner.
The projection with $m=0$ will be denoted by $()_{+}$ and the
corresponding equation will be known as the standard Moyal-Lax representation, while the other two cases will be known as non-standard representations. Let us note the important 
property
\begin{equation}
\lim_{\kappa \rightarrow 0}\;  \Big ( \Lambda \star \Lambda^{'} \Big )_{\geq m} = \Big ( \Lambda \Lambda^{'} \Big)_
{\geq m},
\end{equation}
where the factors on the right hand side are functions on the phase
space (not operators). One of the advantages 
of this method is now obvious, namely, one can go to the Lax representation
of the model in the dispersionless limit in a natural manner
\cite{10}. In  fact, if 
we consider the limit $\kappa \rightarrow 0$, in such a model, it
leads to 
\begin{equation}
{\partial \Lambda_{n}\over \partial t} = \Big \{\Lambda_n,  \Big
(\Lambda_n^{\frac{k}{n}}\Big )_{\geq m}\Big \}. 
\end{equation}
where the bracket on the right hand side is the standard Poisson
bracket. With these definitions, one can construct the conserved charges as
\begin{equation}
H_{k} = Tr {~} \Lambda_{n}^{\frac{k}{n}}, {~}{~~~~} k \neq ln,
\end{equation}
where $k,l$ are integers, prove that different flows commute and can
define Hamiltonian structures in a  straightforward manner \cite{8}. Let us
illustrate these ideas through some examples.

\section*{III. Examples:}
{\bf A.) KdV hierarchy}:

Let us consider the Lax operator 
\begin{equation}
L = p^2 + u(x),
\end{equation}
then, it is straightforward to calculate (remember the projection is
with respect to the star product) 
\begin{equation}
\Big(L^{\frac{3}{2}}\Big)_{+} = p^3 + \frac{3}{2} u\star p
-\frac{2\kappa}{2} u^{(1)}, 
\end{equation}
where $u^{(1)} = \frac{\partial u}{\partial x}$. 

Therefore, the Moyal-Lax equation 
\begin{equation}
\frac{\partial L}{\partial t} = 
\Big \{L,  \Big (L^{\frac{3}{2}}\Big )_{+}\Big \}_{\kappa}, 
\end{equation}
gives
\begin{equation}
\frac{\partial u}{\partial t} = - \Big ( \kappa u^3 + \frac{3}{2} uu^{(1)} \Big ),
\end{equation}
which is the KdV equation and the connection between the parameter of non-commutativity and the central charge 
begins to emerge. However, we will see a more direct relation when we calculate the Hamiltonian structure. 
The conserved quantities can be determined in a straightforward manner and the first few have the forms
\begin{eqnarray}    
H_1 &=& Tr L^{\frac{1}{2}} = \int dx {~} \frac{u}{2}, \nonumber \\
H_2 &=& Tr L^{\frac{3}{2}} = \int dx {~} \frac{u^2}{4}, \nonumber \\
H_3 &=& Tr L^{\frac{5}{2}} = \int dx {~} (4 \kappa  u^{(2)}u + u^3). 
\end{eqnarray}
The commutativity of the flows follows directly from the Moyal-Lax
representation \cite{8}.  

Let us next turn to the question of Hamiltonian structures. First, we define the dual to the Lax operator
in (13) by 
\begin{equation}    
Q = p^{-2}\star q_{-2} + p^{-1}\star q_{-1},
\end{equation}
which allows us to define linear functionals as
\begin{eqnarray}    
F_{Q}(L) &=& Tr {~} LQ = \int dx {~} uq_{-1} , \nonumber \\
F_{V}(L) &=& Tr {~} LV = \int dx {~} uv_{-1} .\
\end{eqnarray}
then, one can define, in a standard manner, the first two Hamiltonian structures of the system as
\begin{eqnarray}    
\Big \{ F_{Q}(L), F_{V}(L) \Big \}_{1} & = & Tr  L\star \Big \{ Q,V\Big \}_{\kappa}, \nonumber \\
\Big \{ F_{Q}(L), F_{V}(L) \Big \}_{2} & = & Tr  \Big  ( ( \Big \{ L,Q \Big \}_{\kappa} )_{+}\star 
(L\star V) + (Q\star L)_{+}\star \Big \{ L,V \Big \}_{\kappa} \Big ) \nonumber \\
&& +\frac{1}{2}\int dx {~} (\int^{x} Res \Big\{Q,L\Big \}_{\kappa})(Res\Big \{V,L \Big \}_{\kappa} ).
\end{eqnarray}
A direct calculations yields
\begin{eqnarray}    
\Big \{ u(x),u(y) \Big \}_{1} &=& 2\frac{\partial }{\partial x} \delta(x-y), \nonumber \\
\Big \{ u(x),u(y) \Big \}_{2} &=& \Big ( u(x)\frac{\partial}{\partial x} + \frac{\partial }{\partial x} u(x)
+2\kappa^2\frac{\partial^3}{\partial x^3} \Big )\delta(x-y),
\end{eqnarray}
which are indeed the two Hamiltonian structures of KdV. Furthermore,
the parameter of non-commutativity, $\kappa$, is 
now seen to be directly related to the central charge of the second
Hamiltonian structure, which is known  to be the Virasoro algebra. 

{\bf B.) Two boson hierarchy}:

Let us next consider the Lax operator 
\begin{equation}    
L = p - J_{0} + p^{-1} \star J_{1}. 
\end{equation}
it is easy to calculate (projection with respect to the star product) 
\begin{equation}    
(L^2)_{\geq 1} = p^2 - 2J_{0}\star p, 
\end{equation}
which, then yields from the nonstandard Moyal-Lax equation
\footnote{Notice that if we take the projection with respect to the
usual product (and not the star product), then, the equation becomes
inconsistent.} 
\begin{equation}    
\frac{\partial L}{\partial t} = \Big \{ (L^2)_{\geq 1} , L \Big \}_{\kappa},
\end{equation}
the two boson equations
\begin{eqnarray}    
\frac{\partial J_{0}}{\partial t} &=& (2J_{0} + J_{0}^2 - 2\kappa J_{0}^{'})^{'},\nonumber \\
\frac{\partial J_{1}}{\partial t} &=& (2J_{0}J_{1} + 2\kappa J_{1}^{'})^{'},
\end{eqnarray}
where the prime denotes derivative with respect to $ x $. Once again there is already a hint of the relation 
between the parameter $\kappa$ and the central charge which we will
see more  explicitly soon. The conserved 
quantities of the  system are defined as 
\begin{equation}    
H_n = Tr L^n,
\end{equation}
and the first few have the explicit forms 
\begin{eqnarray}    
H_1 &=&  \int dx {~} J_1 ,\nonumber \\
H_2 &=& -2\int dx {~} J_{0}J_{1}, \nonumber \\
H_3 &=& 3\int dx {~} (J_{1}^2 + J_{0}^2J_{1} - 2\kappa J_{0}^{'}J_{1} ).
\end{eqnarray}
To study the Hamiltonian structures we define the dual to $L$ as
\begin{equation}    
Q = q_{0} + q_{-1}\star p^{-1},
\end{equation}
so that the linear functionals take the forms
\begin{equation}    
F_{Q}(L) = Tr {~} LQ = \int dx {~} (q_{0}J_{1} - q_{-1}J_{0}).
\end{equation}
The first two Hamiltonian structures can now be defined in a
straightforward manner and have the forms 
\begin{equation}    
\Big \{ F_{Q}(L), F_{V}(L) \Big \}_{1} =  Tr  L\star \Big \{ Q,V\Big
\}_{\kappa}, \nonumber \\ 
\end{equation}
\begin{eqnarray}
&& \Big \{ F_{Q}(L), F_{V}(L) \Big \}_{2}  =  Tr  \Big  ( ( \Big
\{ L,Q \Big \}_{\kappa} )_{+}\star 
(L\star V) + (Q\star L)_{+}\star \Big \{ L,V \Big \}_{\kappa} \Big ) - \nonumber \\
&& \int dx {~} \left(Res\Big \{Q,L\Big \}_{\kappa}Res(L\star V\star p^{-1})
- Res\Big \{V,L\Big \}_{\kappa} 
Res(L\star Q\star p^{-1})\right) \nonumber \\
&& {~~~~~~~~~} {~}{~}+\int dx {~} (\int^{x} Res \Big\{Q,L\Big \}_{\kappa})(Res\Big \{V,L 
\Big \}_{\kappa} ).  
\end{eqnarray}
A straightforward calculation yields
\begin{eqnarray}
\left(\begin{array}{cc}
\Big \{ J_0,J_0 \Big \}_1 & \Big \{ J_0,J_1 \Big \}_1 \\
\Big \{ J_1,J_0 \Big \}_1 & \Big \{ J_1,J_1 \Big \}_1 
\end{array} \right) 
&=& -
\left(\begin{array}{cc}
0 & \partial \\
\partial & 0 
\end{array} \right) \delta(x-y),\nonumber\\
\left(\begin{array}{cc}
\Big \{ J_0,J_0 \Big \}_2 & \Big \{ J_0,J_1 \Big \}_2 \\
\Big \{ J_1,J_0 \Big \}_2 & \Big \{ J_1,J_1 \Big \}_2 
\end{array} \right) 
& =& 
\left(\begin{array}{cc}
2\partial & \partial J_0 -2\kappa \partial^2 \\
J_0\partial +2\kappa\partial^2  & \partial J_1+J_1\partial 
\end{array} \right) \delta(x-y),
\end{eqnarray}
which are the usual Hamiltonian structures of the two boson hierarchy. The second Hamiltonian structure, in 
particular, is the bosonic limit of the $N=2$ twisted  superconformal algebra (one has to 
redefine the basis to make an exact identification) \cite{11} and the relation between $\kappa$ and the central charge 
of the algebra is now explicit.

It is known that the two boson equation reduces to many other integrable models. Without going into details, 
let us note that if we identify 
\begin{equation}
J_0= - \frac{q^{'}}{q}, \quad J_1=\hat q q,
\end{equation}
then the Lax operator in (21) can be rewritten as 
\begin{equation}
L=q^{-1}\star \widetilde{L} \star q,
\end{equation}
where 
\begin{equation}
\widetilde{L} =p + q\star p^{-1}\star \hat q.
\end{equation}
In other words, the two Lax operators $L$ and $\widetilde{L}$ are related through a gauge transformation. It is 
now easy to check that the standard Moyal-Lax equation 
\begin{equation}
\frac{\partial \widetilde{L}}{\partial t} = \Big \{ (\widetilde{L})^2_{+},\widetilde{L} \Big \}_{\kappa},
\end{equation}
leads to the non-linear Schr\"odinger equation while, with the identification $\hat q=q$, the equation 
\begin{equation}
\frac{\partial \widetilde{L}}{\partial t} = \Big \{ (\widetilde{L})^3_{+},\widetilde{L} \Big \}_{\kappa},
\end{equation}
yields the MKdV equation.

\section*{IV. Moyal-Lax representation as a Hamiltonian Equation:} 
The conventional Lax equation (in the standard representation) 
\begin{equation}
\frac{\partial L}{\partial t_{k}} =\Big [ (L^{\frac{k}{n}})_{+},L \Big ],
\end{equation}
resembles a Hamiltonian equation with  $(L^{\frac{k}{n}})_{+}$ reminiscent of the Hamiltonian. However,
such a relation cannot be further quantified in the language of pseudo-differential operators. In contrast, 
we will show now that the Moyal-Lax representation has such a natural interpretation. 

For concreteness, let us consider an arbitrary flow in the KdV hierarchy described by 
\begin{equation}
\frac{\partial L}{\partial t} =\Big \{ L, (L^{\frac{2n+1}{2}} )_{+} \Big \}_{\kappa}. 
\end{equation}
Let us next consider an action of the form 
\begin{equation}
S=\int dt {~} (p\star \dot{x} - (L^{\frac{2n+1}{2}})_{+}).
\end{equation}
The important point to remember, in this, is the fact that $L=L(p,x)$, but does not depend on time
explicitly. Thus, we can think of $(L^{\frac{2n+1}{2}})_{+}$ as the Hamiltonian on the phase space. That this 
is true follows from the Euler-Lagrange equations of the system, namely
\begin{eqnarray}
\dot{x} &=& {~}{~} \frac{\partial 
(L^{\frac{2n+1}{2} })_{+} }{\partial p} {~}{~} ={~} \Big \{ x, 
(L^{\frac{2n+1}{2}})_{+} \Big \}_{\kappa}, \nonumber \\
\dot{p} &=& - \frac{\partial (L^{\frac{2n+1}{2}})_{+}}{\partial x} {~}{~}={~} \Big \{p,
(L^{\frac{2n+1}{2}})_{+} \Big \}_{\kappa}.
\end{eqnarray}
These are indeed Hamiltonian equations with Moyal brackets playing the
role of Poisson brackets, provided
we identify the Hamiltonian of the system with 
$(L^{\frac{2n+1}{2}})_{+}$. It also follows now that, since $L$ is a
function on this phase space, 
\begin{equation}
\frac{\partial L}{\partial t} =\Big \{ L, (L^{\frac{2n+1}{2}} )_{+} \Big \}_{\kappa}. 
\end{equation}
Namely, the Moyal-Lax equation is indeed a Hamiltonian equation with
$(L^{\frac{2n+1}{2}})_{+}$ playing the  
role of the Hamiltonian. Furthermore, the Moyal-Lax equation, as we
have shown, can be derived from an action. Furthermore, although we
have shown this for a standard Moyal-Lax representation, it is clear
that this derivation will go through for nonstandard representations
as well.

\section*{V. Hamiltonian Structures for Dispersionless Systems:}
The Moyal-Lax representation, of course, has the built in advantage that one can go to the dispersionless limit of 
an integrable system by simply taking the limit $ \kappa \rightarrow 0$. While the Lax representations for various 
dispersionless integrable models are known \cite{12}-\cite{14}, the determination of the Hamiltonian structures (at least the second) 
from such a Lax function has remained an open question. The Moyal-Lax representation provides a solution to 
this problem in a natural way. Let us illustrate this with two examples.

First, let us consider the KdV hierarchy, which in the dispersionless
limit, goes over to the Riemann hierarchy \cite{12}. With 
\begin{eqnarray}
L & = & p^2 + u , \quad \quad Q = p^{-2}q_{-2}  + p^{-1}q_{-1}, \nonumber \\
F_{Q}(L) & = & Tr {~} LQ = \int dx {~} uq_{-1},
\end{eqnarray}
we note that the definition of the first two Hamiltonian structures
(19) reduces, in the dispersionless limit, to 
$(\kappa \rightarrow 0)$
\begin{eqnarray}
\Big \{ F_{Q}(L),F_{V}(L) \Big \}_{1} &=& Tr {~} L \Big \{ Q,V \Big \},\nonumber \\
\Big \{ F_{Q}(L),F_{V}(L) \Big \}_{2} &=& Tr {~} \left(( \Big \{ L,Q\Big \})_{+}LV +
(QL)_{+}\Big \{ L,V\Big \}\right) + \nonumber \\
&& \frac{1}{2}\int dx {~} (\int^{x}Res{~}\Big \{Q,L\Big \}){~}(Res{~}\Big \{V,L \Big \}).
\end{eqnarray}
A straightforward calculation leads to 
\begin{eqnarray}
\Big \{ u(x),u(y) \Big \}_{1} &=& 2\frac{\partial}{\partial x} \delta(x-y), \nonumber \\
\Big \{ u(x),u(y) \Big \}_{2} &=& (u(x)+u(y))\frac{\partial}{\partial x}{~}\delta(x-y).
\end{eqnarray}
These are indeed the correct Hamiltonian structures of the Riemann equation and while the first structure 
was already constructed from the Lax function, the construction of the
second structure, from the Lax description, was not known so far \cite{12}.

As a second example, let us consider polytropic gas \cite{13} with $\gamma
= 2$  which can be thought of as the dispersionless 
limit  of the two boson hierarchy. In such a case, the Lax function has the form 
\begin{equation}
L=p+u+vp^{-1}. 
\end{equation}
Defining the dual and the linear functional as 
\begin{eqnarray}
Q &=& q_0+q_{-1}p^{-1}, \nonumber \\
F_{Q}(L) &=& Tr {~} LQ = \int dx {~} (uq_{-1} + vq_0),
\end{eqnarray}
we note that the definition of the first two Hamiltonian structures follows from the dispersionless limit 
$\kappa \rightarrow 0$ of eqs. (30)-(31) to be 
\begin{eqnarray}
&& \Big \{ F_{Q}(L), F_{V}(L) \Big \}_{1} = Tr {~} L\Big \{ Q,V \Big \},\nonumber \\
&& \Big \{ F_{Q}(L), F_{V}(L) \Big \}_{2} = Tr {~} \left(( \Big \{ L,Q\Big \})_{+}LV +
(QL)_{+}\Big \{ L,V\Big \}\right) - \nonumber \\
&&{~}{~} \int {~} dx \left(Res\Big \{Q,L\Big \} Res(LVp^{-1}) -
Res\Big \{V,L\Big \}Res(LQp^{-1})\right)\nonumber \\  
&&{~}{~}\quad \quad +  \frac{1}{2}\int dx {~} (\int^{x}Res{~}\Big \{Q,L\Big \}){~}(Res{~}\Big \{V,L \Big \}).
\end{eqnarray}
A simple calculation yields
\begin{eqnarray}
\left(\begin{array}{cc}
\Big \{ u,u \Big \}_1 & \Big \{ u,v \Big \}_1 \\
\Big \{ v,u\Big \}_1 & \Big \{v,v \Big \}_1 
\end{array} \right) & =& -
\left(\begin{array}{cc}
0 & \partial \\
\partial  & 0 
\end{array} \right) \delta(x-y), \nonumber \\
\left(\begin{array}{cc}
\Big \{ u,u\Big \}_2 & \Big \{ u,v \Big \}_2 \\
\Big \{ v,u \Big \}_2 & \Big \{ v,v \Big \}_2 
\end{array} \right) &=& 
\left(\begin{array}{cc}
2\partial & \partial u \\
u \partial & \partial v + v\partial  
\end{array} \right) \delta(x-y).
\end{eqnarray}
These are indeed the correct Hamiltonian structures of this model. We would like to emphasize that it was
not known so far how to derive the second Hamiltonian structure from
the Lax description of the system \cite{13}. We 
would also like  to note here that the Lax function for the polytropic
gas with arbitrary, large $\gamma$ \cite{13} is 
highly constrained. Consequently the formulae in (48) need to be
modified further (even for the first Hamiltonian structure)  
and we have not yet analyzed this question.

To summarize, we have studied the properties of Moyal-Lax representation systematically in this letter. In 
addition to the fact that they allow a smooth passage to the dispersionless models, we have shown that the 
parameter of non-commutativity is related to the central charge of
the second Hamiltonian structure of the system. We have  
shown that the Moyal-Lax equation can be interpreted as a Hamiltonian equation and can be derived from an action. 
We have also shown how the Moyal-Lax description leads in a natural manner, in the dispersionless limit, to the 
Hamiltonian structures of dispersionless integrable models which has been an open question for sometime. 
In many ways, this alternate description of integrable systems seems
more desirable. Properties of the Moyal-Lax representation for
supersymmetric integrable systems will be described separately \cite{15}.

\section*{Acknowledgments}
One of us (A.D) would like to thank the organizers of the 37th Karpacz Winter School as well as the members of the 
Institute of Theoretical Physics, Wroc\l aw fort hospitality, where this work was done. This work was supported 
in part  by US DOE Grant No. DE-FG 02-91ER40685 and by NSF-INT-0089589.

\noindent {\it Note added:} We would like to thank C. Zachos for
pointing out to us that the star product is really due to Groenewold
\cite{3'}. We would also like to thank I. Strachan as well as the
referee for bringing \cite{5'}, which has some overlap with our work, to
our attention.


\begin{thebibliography}{99}
\bibitem{1} G. B. Whitham, {\it Linear and Nonlinear Waves}, John
Wiley, 1974; L. D. Faddeev and L. A. Takhtajan, {\it Hamiltonian Methods and the Theory of Solitons}, Springer, 1987; 
A. Das, {\it Integrable Models}, World Scientific, 1989;
M. Blaszak {\it Multi-Hamiltonian Theory of Dynamical System} Springer Verlag 1998.
\bibitem{2} P. Mathieu, 
J. Math. Phys. {\bf 29} (1988) 2499; W. Oevel and Z. Popowicz
Comm. Math. Phys. {\bf 139} (1991) 441; J. C. Brunelli and  A. Das, Int. Jour. of Modern Physic {\bf
10} No.32 (1995) 4563.  
\bibitem{3} N. Seiberg and E. Witten, {\it String theory and
non-commutative geometry}, hep-th/9908142.
\bibitem{3'} H. Groenewold, Physica {\bf 12} (1946) 405.
\bibitem{4} J. E. Moyal, Proc. Cambridge Phil. Soc. {\bf 45} (1949) 99.
\bibitem{5} A. F. Dimakis and F. M\"{u}ller-Hoissen,
Rep. Math. Phys. {\bf 46} (2000) 203;  {\it Non-Commutative 
Korteweg--de-Vries equation}, hep-th 0007074.
\bibitem{5'} B. Kupershmidt, Lett. Math. Phys. {\bf 20} (1990) 19.
\bibitem{6} I. Strachan, Phys. Lett {\bf B283} (1992) 63; J. Phys. {\bf
A29} (1996) 6117.
\bibitem{6'} H. Garcia-Campean and J. Plebanski, Phys. Lett. {\bf A
234} (1997) 85; C. Zachos, D. Fairlie and T. Curtright, {\it Matrix
membranes and integrability}, in {\it Supersymmetry and Integrable
Models}, ed. H. Aratyn et al, Springer LNP 502, Germany (1998).
\bibitem{7} T. Koikawa, Phys. Lett {\bf A256}
(1999) 284; Prog. Theor. Phys. {\bf 102} (1999) 29; {\it Soliton
equation extracted  from the noncommutativite 
zero-curvature equation}, hep -th 0101067. 
\bibitem{8} A. Das and W. J. Huang, J. Math. Phys. {\bf 33} (1992) 2487.
\bibitem{9} K. Takesaki {\it Nonabelian KP hierarchy with Moyal
algebraic coeffcients},  hep-th/9305169,
Kyoto University KUCP - 0062/93.
\bibitem{10} D. Lebedev and Y. I. Manin, Phys. Lett {\bf 74A} (1979)
154; V. E. Zakharov, Physica {\bf 3D} (1981) 
193; Y. Kodama and J.Gibbons, Phys. Lett. {\bf 135A} (1989) 167;
K. Takesaki and T. Takebe, Rev.Math.Phys. 
{\bf 7} (1995) 57.
\bibitem{11} A. Das and S. Roy, Mod. Phys. Lett. {\bf A11} (1996)
1317.
\bibitem{12} J. C. Brunelli, Rev. Math. Phys. {\bf 8} (1996) 1041.
\bibitem{13} J. C. Brunelli and A. Das, Phys. Lett. {\bf A235} (1997)
597; J. C. Brunelli and A. Das, Phys. Lett. {\bf B426} (1998) 57.
\bibitem{14} J. Barcelos-Neto, A. Constandache and A. Das,
Phys. Lett. {\bf A268} (2000) 342; A. Das and Z. Popowicz,
Phys. Lett. {\bf A272} (2000) 65.
\bibitem{15} A. Das and Z. Popowicz, {\it Supersymmetric Moyal-Lax
Representation}, hep-th/0104191.
\end{thebibliography}
\end{document}